\documentclass[english]{article}
\usepackage{mathptmx}
\usepackage[T1]{fontenc}
\usepackage[latin9]{inputenc}
\usepackage[a4paper]{geometry}
\usepackage{array}
\usepackage{textcomp}
\usepackage{amstext}
\usepackage{graphicx}
\usepackage{setspace}

\makeatletter

\providecommand{\tabularnewline}{\\}

\makeatother

\usepackage{babel}
\begin{document}
\begin{onehalfspace}

\title{Effect of phosphous-doped upon the electronic structures of single
wall carbon nanotubes}
\end{onehalfspace}

\begin{onehalfspace}

\author{AQing Chen, QingYi Shao%
\thanks{corresponding author(email:qyshao@163.com)%
}, ZhiCheng Lin}
\end{onehalfspace}

\maketitle
\begin{onehalfspace}
School of Physics \& Telecommunication Engineering, South China Normal
University, GuangZhou 510006, China
\end{onehalfspace}
\begin{abstract}
\begin{onehalfspace}
We studied Phosphorus-doped single wall carbon nanotubes (SWCNT) by
using the First-Principle method based on Density Function Theory
(DFT). The formation energy, total energy, band structure, geometry
structure and density of states were calculated. We have found that
the formation energy of the P-doped single carbon nanotube increases
with its diameter. The total energy of carbon nanotubes in the same
diameter decreases with the increasing doping concentration. The effects
of impurity position on the impurity level were discussed in this
paper. It is illustrated that the position of impurity level may depend
on the C-P-C bond angle. According to the results, it is feasible
to substitute a phosphorus atom for a carbon atom in SWCNT. It is
also found that P-doped carbon nanotubes are n type semiconductors.\end{onehalfspace}

\end{abstract}
\begin{onehalfspace}
\end{onehalfspace}

\begin{onehalfspace}

\section{Introduction}
\end{onehalfspace}

\begin{onehalfspace}
Since the carbon nanotubes were discovered \cite{iijima_helical_1991},
they have attracted the attention of numerous research groups because
of their outstanding mechanical and electronic properties. A single-wall
carbon nanotube (SWCNT) can be described as a graphite sheet rolled
into a cylindrical shape so that the structure is of one dimension
\cite{jose-yacaman_catalytic_1993} with a diameter of about 0.7-10.0
nm. Carbon nanotubes are regarded as the ultimate fiber with regard
to its strength in the direction of the nanotubes axis \cite{overney_structural_1993}.
Carbon nanotubes are regarded as the ultimate fiber with regard to
its strength in the direction of the nanotubes axis , and its electronic
structure can be either metallic or semiconducting depending on its
diameter and chirality \cite{dresselhaus_physics_1995}. Carbon nanotubes
have been studied in a wide range of areas, including field emitter
for flat panel displays, microelectronics devices, chemic sensors
\cite{cao_electromechanical_2003} press sensors\cite{grow_piezoresistance_2005},
etc. Doped carbon nanotubes have played a vital role in the electronic
devices and nano-sensors fabricated with carbon nanotubes. There are
many investigations on doped carbon nanotubes in theory such as boron-doped
carbon nanotubes and N-doped carbon nanotubes. It is known that B
impurities are injected in carbon nanotubes with small diameter easier
than those with big diameter. The ionization potential (IP) of B-doped
zigzag (10,0) carbon nanotubes is about 0.2 eV. The work function
of N-doped carbon nanotubes is lower than that of pristine carbon
nanotubes. Otherwise, work function and ionization potential of N-doped
carbon nanotubes vary with the position of N impurities in carbon
nanotubes. Both decrease on energy gap of semiconducting carbon nanotubes
and increase in local density of states on Fermi energy level owing
to the presence of impurities. 

It is because carbon nanotubes behave as a metal or as a semiconductor
depending on the chirality and diameter that it is hard to attain
homogeneous carbon nanotubes in experiment. The doped carbon nanotubes
will be P type or N type semiconductor, just as diamond film which
is doped with phosphorus is N type semiconductor. P-doped diamond
film has been synthesized with microwave inductive plasma chemic vapor
deposition by Chen et al.\cite{chen_electronic_2001}. They found
that the filed emission current of P-doped diamond film has been increased
sharply more than that of B-doped diamond film. Cruzsilva et al. \cite{cruz-silva_heterodoped_2008}
used a solution of ferrocene, triphenylphosphine, and benzylamine
in conjunction with spray pyrolysis to synthesize phosphorus-nitrogen
doped multiwall carbon nanotubes forming hetero-doped carbon nanotubes
and did some theoretic calculation. However, they did not study the
electronic structure further, such as how both the concentration of
P impurity and the position of P atom in SWCNT affect the electronic
structure, while we discuss this point in detail, right now. 

As a new material of nanotechnology, it is necessary that carbon nanotubes
should be doped with impurities. It is of a vital importance to study
the effects of concentration on the total energy of carbon nanotubes
and to predict the feasibility of attaining carbon nanotubes with
different impurity concentrations. Different positions of P impurity
in carbon nanotubes have different effects on the electronic structure
of carbon nanotubes. So it is inevitable to study the doped form of
P impurity and the effects of P impurity position on carbon nanotubes.
It maybe provide some important theory for investigating the device
applied to nano-integrate circuit in the future. 

As far as I know, there is little theoretic calculation on P-doped
SWCNT. But it is only to study the energy band structure and total
energy without studying the effects of position of P impurity on the
electronic structure. Considering the above important facts, P-doped
SWCNT will be introduced in this letter, and the first principle calculation
based density of function theory is carried out to explore the basic
electronic structure characteris- tic of P-doped SWCNT by attaining
the formation energy change of SWCNT after P presence and analyzing
the effects of the position of P in SWCNT. There are three aspects
to be calculated: the geometry structure and total energy of P-doped
SWCNT with different diameters, the total energy of P-doped SWCNT
with vari- ous impurity concentrations, and analysis of the interre-
lated factor of position of impurities level induced by the position
of P impurity. The formation energy varies with the diameter of P-doped
SWCNT. 
\end{onehalfspace}

\begin{onehalfspace}

\section{The model and method of calculation }
\end{onehalfspace}

\begin{onehalfspace}
All theoretical calculations in this letter are completed by Castep
package \cite{segall_first-principles_2002} which is a state-of-the-art
quantum mechanics-based program designed specifically for solid-state
materials science and employs the density functional theory plane-wave
pseudopotential method. It allows us to perform first-principles quantum
mechanics calculations to explore the properties of crystals and surfaces
in materials such as semiconductors, ceramics, metals, minerals, and
zeolites, and it is available from Accelry Inc. It is the more precise
calculation method to explore the electronic structure currently \cite{watanabe_formation_2001}.
Many calculations in published papers \cite{yan_density_2007}, \cite{zhao_electronic_2004}
have been completed by it. In this letter, we focus on zigzag tubes
and arm- chair tube including (5,0), (6,0), (7,0), (8,0), (9,0) and
(10,0) type tube, in whose cell there are 20, 24, 28, 32, 36, and
40 atoms, respectively. We substitute a C atom in carbon framework
with a P atom. In order to study the effects of the impurity concentration
upon SWCNT, we construct PC$_{\text{39}}$, P$_{\text{2}}$C$_{\text{38}}$,
and P$_{\text{3}}$C$_{\text{37}}$ tubes, respectively. The position
of impurities is different. All of the models have been calculated
by the First Principle Theory based on DFT. We have used the general-gradient
potential approximation (GGA) \cite{perdew_generalized_1997} of PBE
for exchange and correlation functional. GGA provides a better overall
description of the electronic subsystem than the LDA func- tional\textquoteright{}s
in the framework of DFT because the LDA description tends to overbind
atoms, so that the bond lengths and the cell volume are usually underestimated
by a few percent and the bulk modulus is correspond- ingly overestimated.
The calculations are expended us- ing plane waves basis with a cutoff
energy of 470 eV, and we also use Normal-conserving pseudopotential
\cite{troullier_efficient_1991}, \cite{kleinman_efficacious_1982}
to perform the interaction between ions and electrons. Geometry optimization
is performed with convergence toleration of 2$\times$10$^{\text{-5}}$
eV/atom. The pressure on every atom is less than 0.05 eV/Å and stress
is less than 0.1 GPa. We take 1$\times$1$\times$6 k points in the
first Brillouin zone. 
\end{onehalfspace}

\begin{onehalfspace}

\section{Results}
\end{onehalfspace}

\begin{onehalfspace}

\subsection{The optimized geometry structure and total energy }
\end{onehalfspace}

\begin{onehalfspace}
The geometry structure of P-doped SWCNTs with different diameters
has been optimized. Figure 1 illustrates the structure of (5,0) P-doped
SWCNT clearly. We can see that a C atom in hexagon of a carbon nanotube
frame is substituted with a P atom. Because the P atom radius is bigger
than C atom radius, the regular hexagon frame is distorted. The length
of C-P bond is 1.79 Å. The angles of C-P-C is varied from 94.042\textdegree{}-100.348\textdegree{}
just as described in Figure 1, which agrees with the results reported
by Cruz-Silva \cite{cruz-silva_heterodoped_2008}. P atom has been
pushed out from the graphene layer in order to form a stable framework.
\begin{figure}
\begin{centering}
\includegraphics[width=6.5cm,height=6.5cm]{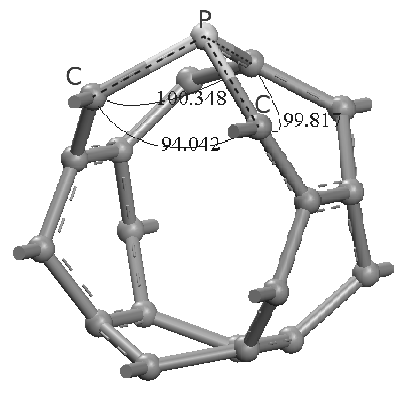}
\par\end{centering}

\caption{The structure of (5,0) type P-doped SWCNT. }
\end{figure}

The formation energy{[}9{]} of the substitutional P impurity by comparing
the total energy to that of a pristine SWCNT has been calculated to
discuss the stability of a P-doped SWCNT. The formation energy can
be defined by the following formula: 

\[
\triangle E=E\left(PC_{n-1}\right)-\frac{n-1}{n}E\left(C_{n}\right)-E\left(P\right)
\]

Here, $E\left(PC_{n-1}\right)$and $E\left(C_{n}\right)$ represent
the total energies of a P-doped and pristine SWCNT, respectively,
and $E\left(P\right)$ represents the energy per atom in phosphorus.
Because $E\left(P\right)$ is constant, the above formula can be rewritten
as 

\[
\delta=\triangle E+E\left(P\right)=E\left(PC_{n-1}\right)-\frac{n-1}{n}E\left(C_{n}\right)
\]

Table 1 indicates the relationship between the diameter of P-doped
SWCNT and $\delta$. It is noteworthy that $\delta$ increases with
diameter in both zigzag and armchair tubes indicating that narrower
tubes are more favorable for phosphorus doping, which is similar to
the results of B-doped SWCNT reported by Koretsune et al. \cite{koretsune_electronic_2008}.
According to their theory, we think that in order to stretch the P-C
bonds, the carbon atoms should be pushed or the P atom should be moved
higher the graphite layer accompanied by symmetry breaking, when graphite
layer is doped substitutionally with P atom. For the tubes with big
curvature, P atom can be injected when the tubes are distorted slightly.
However, for the tubes with small curvature they need to be distorted
intensely. But we think that the important reason is the form of C-P
bond. When graphite layer is doped with P atoms, P atoms interact
with the nearby C atoms exhibiting sp2 and sp3 hybridization. But
there is more sp3 hybridization in tubes with narrower diameter accompanied
with lower energy. So, the P-doped SWCNT with narrow diameters is
easily synthesized. We have calculated the total energy to discuss
the sta- bility of P-doped SWCNT with different impurity concentrations.
The relationship between impurity concentration and total energy is
shown in Table 2. The total energy of doped SWCNT decreases when the
impurity concentration increases, which indicates that it is easier
to synthesize doped carbon nanotubes if the impurity concentration
becomes higher.
\end{onehalfspace}

\begin{table}
\caption{$\delta$ for different diameters }

\medskip{}

\centering{}%
\begin{tabular}{>{\centering}p{1.5in}>{\centering}p{1.5in}}
\hline 
Diameter (Å)  & $\delta$ (eV) \tabularnewline
\hline 
3.91(5,0)  & \textminus{}179.79413\tabularnewline
4.70(6,0)  & \textminus{}179.61113 \tabularnewline
5.48(7,0)  & \textminus{}179.08833 \tabularnewline
6.62(8,0)  & \textminus{}178.99062\tabularnewline
7.05(9,0)  & \textminus{}178.69707 \tabularnewline
7.83(10,0)  & \textminus{}178.4553\tabularnewline
\hline 
\end{tabular}
\end{table}

\begin{onehalfspace}
According to Table 2, the total energy of P-doped SWCNT is lower than
that of pristine SWCNT, which suggests that P-doped SWCNT is more
stable than pristine SWCNT. So it is likely possible to synthesize
P-doped SWCNT with substitutional doping in experiment. 
\begin{table}
\begin{centering}
\caption{The total energy of SWCNT with different P impurity concentration }

\par\end{centering}

\medskip{}

\centering{}%
\begin{tabular}{>{\centering}p{1.5in}>{\centering}p{1.5in}}
\hline 
Type & $E\left(eV\right)$\tabularnewline
\hline 
C$_{20}$ & \textminus{}3101.13 \tabularnewline
PC$_{19}$ & \textminus{}3125.86 \tabularnewline
C$_{24}$ & \textminus{}3724.84 \tabularnewline
PC$_{23}$  & \textminus{}3749.25 \tabularnewline
C$_{28}$  & \textminus{}4348.32 \tabularnewline
PC$_{27}$  & \textminus{}4372.11 \tabularnewline
C$_{40}$  & \textminus{}6217.37 \tabularnewline
PC$_{39}$  & \textminus{}6240.39 \tabularnewline
P$_{2}$C$_{38}$  & \textminus{}6264.36 \tabularnewline
P$_{3}$C$_{37}$ & \textminus{}6287.33\tabularnewline
\hline 
\end{tabular}
\end{table}

\end{onehalfspace}

\begin{onehalfspace}

\subsection{The band structure and density of state }
\end{onehalfspace}

\begin{onehalfspace}
In order to study electrical characteristics of P-doped carbon nanotubes
well, both band structure and density of states of (5,0), (7,0), (8.0)
and (10,0) tubes, respectively are calculated. Figure2 shows the band
structure of (5,0), (7,0), (8,0) and (10,0) P-doped SWCNT, respectively.

\begin{figure}
\begin{centering}
\includegraphics[width=7cm,height=6cm]{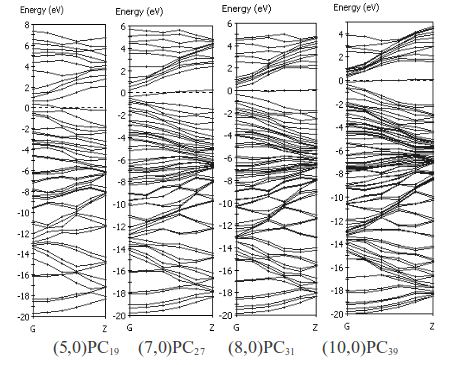}
\par\end{centering}

{\footnotesize \caption{The band structure of (5,0)PC$_{19}$ (a), (7,0)PC$_{27}$ (b), (8,0)PC$_{31}$
(c), and (10,0)PC$_{39}$ (d) type P-doped SWCNT, respectively. }
}
\end{figure}

According to Figure 2, the presence of P introduces an impurity energy
level near the Fermi energy level, while impurity energy levels of
(5,0)PC$_{\text{19}}$, (7,0)PC$_{\text{27}}$, and (10,0)PC$_{\text{39}}$,
respectively are closer to valence band top than to conduction band
bottom across the Fermi energy level. But the impurity level of (8,0)PC$_{\text{31}}$
crosses the Fermi energy in the middle of the gap between valence
band top and conduction band bottom. Just as described in Figures
3(c) and 3(d), the electrons from \textit{p} orbit belong to P atom
without becoming bond. So they mainly contribute to the DOS near Fermi
energy level. Each carbon atom has six electrons which occupy 1\textit{s}$^{\text{2}}$,
2\textit{s}$^{\text{2}}$ and 2\textit{p}$^{\text{2}}$ atomic orbitals.
The 2\textit{s} orbital and two 2\textit{p} orbitals are hybridized
forming three \textit{sp}$^{2}$ hybridized orbitals. These hybrid
orbitals form $\sigma$ bonds. Each carbon atom has an excess electron
which has been hybridized but forms a $\pi$ orbital perpendicular
to the graphite sheet plane. According to Figure 3(d), although there
are numerous electrons from p orbital to contribute to valence band
and conduction band, there are also a few electrons from \textit{s}
orbital to contribute to valence band and conduction band. Besides,
there are few electrons from \textit{s} orbital near Fermi energy
level contributing to valence band and conduction band {[}see Figure
3(b){]}. According to the PDOS of P-doped SWCNT, however, just as
described in Figure 3(d), there are a few electrons from \textit{s}
orbital near Fermi energy level. Therefore the electrons from \textit{s}
orbital are mainly from P atom. By comparing Figures 3(a) with 3(b),
it is obvious that the peak in Fermi energy level mainly results from
the impurity levels of P atoms. Figure 3(c) reveals that some peaks
locating between $\mbox{-}$4 and $-$1 eV are mainly due to electrons
from \textit{p} orbitals of P impurities. But there are also a few
electrons from \textit{s} orbitals for those peaks. There is a mixing
of \textit{p} and \textit{s} atomic orbitals at the same level, which
leads to that \textit{p} orbital and \textit{s} orbital are hybridized.
The peak above Fermi energy level is also mainly due to the result
that \textit{p} orbital and \textit{s} orbital are hybridized. 

\begin{figure}
\begin{centering}
\includegraphics[width=13cm,height=10cm]{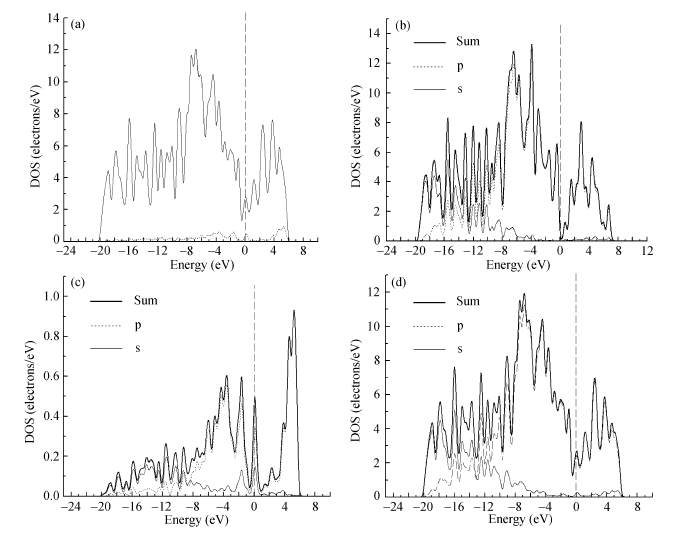}
\par\end{centering}

\caption{Total density of states (TDOS) of (a) P-doped (7,0) SWCNT and TDOS
of P impurities described by the dash line. Partial density of states
(PDOS) of (b) pristine (7,0) SWCNT, (c) P impurities and (d) P-doped
(7,0) SWCNT. The Fermi level is at 0 eV. }

\end{figure}

The band structure of pristine (10,0) SWCNT and P-doped SWCNT is shown
in Figures 4(a)-(d). Pristine zigzag (10,0) carbon nanotubes have
the D$_{10h}$ symmetry \cite{li_effects_2007}, \cite{pan_ab_2004},
\cite{saito_physical_1998}. Figure 4(a) shows the band structure
of pristine (10,0) carbon nanotubes, from which we can see that zigzag
(10,0) carbon nanotube is direct-gap semiconductor and that the energy
bands show a large doubly degeneracy at the zone boundary. The band
gap is 0.57 eV approximately, which is similar to that reported in
some paper{[}14{]}. Valence band top and conduction band bottom correspond
to $\pi$ energy bonding band and $\pi$$^{\text{*}}$energy anti-bonding
band respectively. Figures 4(b)-(d) are the energy band structures
of PC$_{39}$, P$_{2}$C$_{38}$ and P$_{3}$C$_{37}$, respectively
and there are impurity energy levels near Fermi energy level. Figure
4(b) shows clearly that impurity energy levels cross the Fermi energy
level and the impurity energy level is closer to valence band top
than to conduction band bottom. It also shows that Fermi energy is
closer to valence band top. In Figure 4(c) we can see that there are
two impurity energy levels. One crosses Fermi energy but the other
is under Fermi energy level. There are three impurity energy levels
shown in Figure 4(d). One is between Fermi energy level and conduction
band bottom, the other is across Fermi energy and the last one is
between Fermi level and valence band top. 
\begin{figure}
\begin{centering}
\includegraphics[width=11cm,height=8cm]{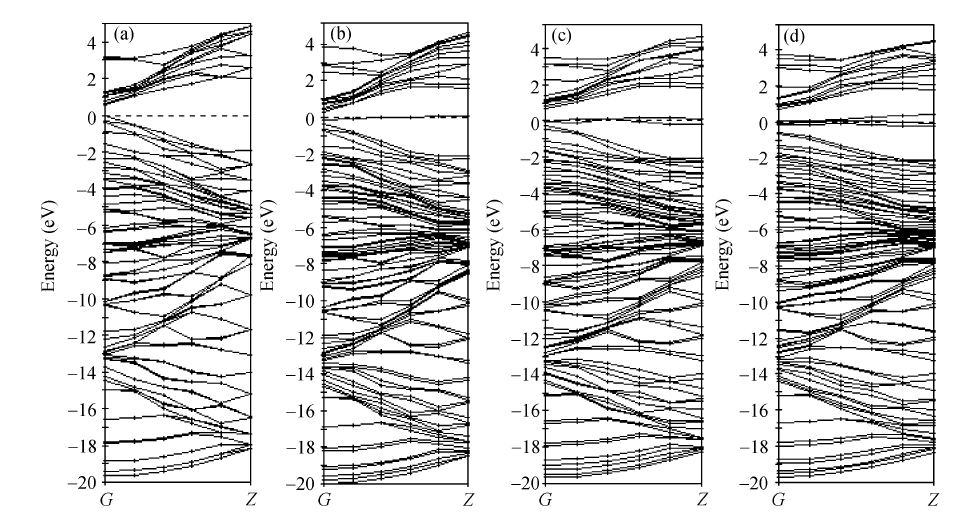}
\par\end{centering}

\caption{Energy band structures of pristine zigzag (10,0) SWCNT and P-doped
(10,0) SWCNT. (a) C$_{40}$; (b) PC$_{39}$; (c) P$_{2}$C$_{38}$;
(d) P$_{3}$C$_{37}$. }
\end{figure}
\begin{figure}
\begin{centering}
\includegraphics{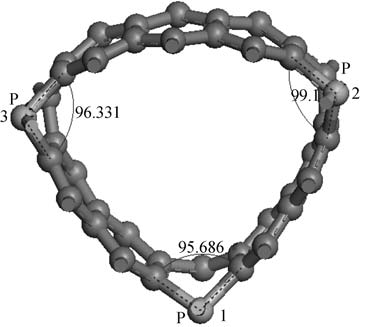}
\par\end{centering}

\caption{The geometry structure of P$_{3}$C$_{37}$SWCNT. }
\end{figure}
\begin{figure}
\begin{centering}
\includegraphics{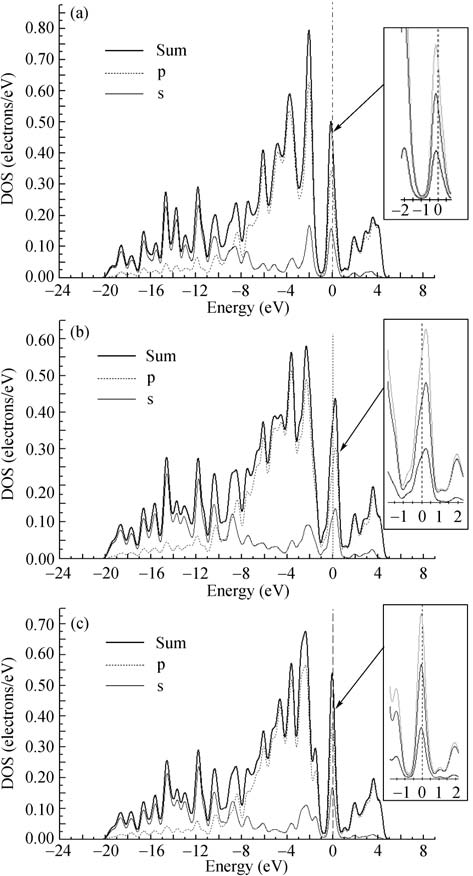}
\par\end{centering}

\caption{PDOS of (a) No. 1 P, (b) No. 2 P and (c) No. 3 P atom, respec- tively.
The right inset shows the sharp peaks near the Fermi energy level. }
\end{figure}

Figure 5 shows the structure of P$_{3}$C$_{37}$ SWCNT and the different
positions of P atom. The DOS of No. 1, No. 2 and No. 3 atoms are shown
in Figures 6(a)-(c), respectively. The peak located at the Fermi energy
level left is an impurity peak just as Figure 6(a) shows. It illustrates
that there is an impurity energy level under the Fermi energy level.
The reason for that is the C-P-C bond angle with 95.686\textdegree{}
which is close to the typical angle of \textit{sp}$^{3}$ hybridization.
It suggests that there may be \textit{sp}$^{3}$ hybridization between
C atoms and P atoms to lead to $\sigma$ bond. So an impurity energy
level is introduced between Fermi energy level and valence band top.
There is a sharp peak at the Fermi energy level right shown in Figure
6(b) which corresponds to the DOS of No. 2 P atom. The C-P-C angle
composed with No. 2 P atom and C atoms is 99.1\textdegree{} (see Figure
5), so there may be no \textit{sp}$^{3}$ hybridization between No.
2 P atom and nearby C atoms. The P atom has one valence electron more
than C atom. Therefore an impurity energy level is generated between
Fermi energy and conduction band bottom. The C-P-C angle composed
with No. 3 P atom and C atoms is among the angle composed with No.
1 P atom and C atoms and the angle composed with No. 2 P atom and
C atoms. Therefore, the No. 3 P impurity energy level is located between
No. 1 P and No. 2 P impurity energy level, which is described well
in Figure 4(d). Other P-doped SWCNTs have the same characteristics
as (10,0) P-doped SWCNT. According to Figure 6, it is noticeable that
the peaks of P impurity near Fermi energy mainly result from that
the \textit{s} and \textit{p} atomic orbitals of P atom are hybridized. 
\end{onehalfspace}

\begin{onehalfspace}

\section{Conclusion}
\end{onehalfspace}

\begin{onehalfspace}
We have studied the basic electronic structure of P-doped SWCNT by
the First Principle Theory based DFT. The formation energy of P-doped
SWCNT is related to the diameters. The narrower diameters are, the
lower formation energy is. Therefore, the system gets more stable.
It can be concluded that the total energy of P-doped SWCNT is lower
than that of pristine SWCNT, and the total energy will get lower with
the impurity concentration increase by analyzing the total energy
of P-doped carbon nanotubes. It is feasible to substitute a carbon
atom with a phosphorus atom in SWCNT in terms of theory. By analyzing
of DOS and PDOS, we can see that the DOS near Fermi energy level attribute
to the electronics from the p orbital and the impurity energy level
may be affected by the C-P-C angle and the sp hybridization leads
to the impurity peak, so the presence of P atom affects the physic
property of SWCNT in a certain sense. Doping is important for chemic
and physic performance of SWCNT as the promising nano-material in
the future. The P-doped carbon nanotubes enrich the material for fabricating
nano-device. The other significance of this paper is that it may provide
some theory for studying the device applied to nano-integrated circuits
in the future. 

\bibliographystyle{unsrt}
\bibliography{cite1}
\end{onehalfspace}

\end{document}